\documentclass[a4paper,fleqn]{cas-sc}

\usepackage[numbers,sort&compress]{natbib}
\usepackage{amsmath,amssymb}
\usepackage{array}
\usepackage{booktabs}
\usepackage{graphicx}
\usepackage{longtable}
\usepackage{tabularx}
\usepackage{xurl}
\usepackage{flafter}
\newsavebox{\fitgraphicbox}
\newcommand{\fitgraphic}[1]{\sbox{\fitgraphicbox}{\includegraphics{#1}}\ifdim\wd\fitgraphicbox>\columnwidth\includegraphics[width=\columnwidth]{#1}\else\usebox{\fitgraphicbox}\fi}
\newcommand{\tablecontentfont}{\footnotesize}
\AtBeginEnvironment{tabular}{\tablecontentfont}
\AtBeginEnvironment{tabularx}{\tablecontentfont}
\AtBeginEnvironment{longtable}{\tablecontentfont}
\hypersetup{hypertexnames=false}
\renewcommand{\printorcid}{}
\begin{document}

\let\WriteBookmarks\relax
\shorttitle{ }
\shortauthors{Zhu et al.}

\title[mode=title]{Virtual Transmission Lines from  Data Center Workload Migration: Coordinated Multi-Year Transmission Expansion Planning}

\author[1]{Mengxin Zhu}
\ead{mxzhu@sjtu.edu.cn}

\author[1]{Qian Kai}
\ead{kaiqian02@sjtu.edu.cn}

\author[1]{Yadong Liu}
\ead{lyd@sjtu.edu.cn}

\author[1]{Yingjie Yan}
\cormark[1]
\ead{yanyingjie@sjtu.edu.cn}

\author[1]{Haoyu Zhao}
\ead{zhaohaoyu@sjtu.edu.cn}

\affiliation[1]{organization={Shanghai Jiao Tong University},
            city={Shanghai},
            country={China}}

\cortext[1]{Corresponding author: Yingjie Yan}

\begin{abstract}
The rapid growth of electricity demand from AI data centers is creating a dual mismatch in power systems: a temporal mismatch between fast-rising demand and slow transmission expansion, and a spatial mismatch between concentrated data-center loads and geographically dispersed renewable resources. These mismatches can exacerbate transmission congestion, resulting in unserved data-center load and renewable-energy curtailment. This study models geographic workload migration as a virtual transmission line (VTL) that changes nodal data-center demand without transferring physical electricity. A mixed-integer linear programming model is then developed to co-optimize multi-year candidate-line construction and workload migration, subject to construction delays, renewable-energy curtailment, unserved data-center load, workload-balance requirements, communication-bandwidth constraints, and link-latency limits. The model is evaluated over an eight-year planning horizon on a modified IEEE 14-bus system. The results show that coordinating VTL with physical transmission expansion reduces total system cost, unserved data-center load, and renewable-energy curtailment while alleviating transmission congestion. VTL-aware planning also requires one fewer candidate line than build-only planning. However, VTL alone cannot meet system demand and does not eliminate the need for physical transmission reinforcement. Sensitivity analysis indicates that VTL provides the greatest value when transfer capacity is tight, construction costs are high, load growth is rapid, and data-center flexibility is greater. Additional validation on the IEEE 118-bus system over a five-year planning horizon shows that coordinated planning retains its cost and reliability benefits while reducing physical expansion requirements.
\end{abstract}

\begin{keywords}
data center \sep virtual transmission line \sep transmission expansion planning \sep workload migration \sep renewable curtailment \sep mixed-integer linear programming
\end{keywords}

\maketitle

\section*{Nomenclature}
\label{sec:nomenclature}
\begingroup
\renewcommand{\arraystretch}{0.76}
\setlength{\tabcolsep}{1.5pt}
\begin{longtable}{@{}p{0.18\textwidth}p{0.30\textwidth}p{0.18\textwidth}p{0.30\textwidth}@{}}
\toprule
Symbol & Meaning & Symbol & Meaning \\
\midrule
\endfirsthead
\multicolumn{4}{l}{\textit{Nomenclature continued}}\\
\toprule
Symbol & Meaning & Symbol & Meaning \\
\midrule
\endhead
\midrule
\multicolumn{4}{r}{\textit{Continued on next page}}\\
\endfoot
\bottomrule
\endlastfoot
$\mathcal{Y}$ & Set of planning years & $y$ & Planning-year index \\
$\tau$ & Construction-start year index & $\mathcal{S}$ & Set of representative days \\
$s$ & Representative-day index & $\mathcal{T}$ & Set of intra-day intervals \\
$t$ & Intra-day interval index & $\mathcal{N}$ & Set of power-system buses \\
$n$ & Power-system bus index & $\mathcal{E}$ & Set of existing lines \\
$e$ & Existing-line index & $\mathcal{C}$ & Set of candidate lines \\
$c$ & Candidate-line index & $\mathcal{G}$ & Set of conventional generators \\
$g$ & Conventional-generator index & $\mathcal{R}$ & Set of renewable generators \\
$r$ & Renewable-generator index & $\mathcal{D}$ & Set of data centers \\
$a$ & Origin data-center index & $b$ & Destination data-center index \\
$\mathcal{K}$ & Set of directed communication links & $\mathcal{G}_n$ & Conventional generators at bus $n$ \\
$\mathcal{R}_n$ & Renewable generators at bus $n$ & $\mathcal{D}_n$ & Data centers at bus $n$ \\
$q$ & Generic transmission-line index & $i(q)$ & Sending bus of line $q$ \\
$j(q)$ & Receiving bus of line $q$ & $n_0$ & DCPF reference bus \\

$P^G_{g,y,s,t}$ & Conventional-generator output & $P^R_{r,y,s,t}$ & Renewable power delivered to the grid \\
$P^{curt}_{r,y,s,t}$ & Curtailed renewable power & $P^{shed}_{a,y,s,t}$ & Unserved data-center workload \\
$P^{exec}_{a,y,s,t}$ & Workload executed at data center $a$ & $M_{a,b,y,s,t}$ & Workload migrated from $a$ to $b$ \\
$F^E_{e,y,s,t}$ & Active-power flow on existing line $e$ & $F^C_{c,y,s,t}$ & Active-power flow on candidate line $c$ \\
$F_{q,y,s,t}$ & Active-power flow on generic line $q$ & $\theta_{n,y,s,t}$ & Voltage phase angle at bus $n$ \\
$x_{c,y}$ & Construction-start indicator & $z_{c,y}$ & In-service indicator \\
$P^L_{n,y,s,t}$ & Conventional demand at bus $n$ & $\overline P^R_{r,y,s,t}$ & Renewable power available before curtailment \\
$\overline P^R_{r,0}$ & Initial renewable capacity & $\underline P^G_g$ & Minimum conventional-generator output \\
$\overline P^G_g$ & Maximum conventional-generator output & $X_e$ & Reactance of existing line $e$ \\
$X_c$ & Reactance of candidate line $c$ & $b_e$ & Susceptance coefficient of existing line $e$ \\
$b_c$ & Susceptance coefficient of candidate line $c$ & $\overline F_e$ & Flow limit of existing line $e$ \\
$\overline F_c$ & Flow limit of candidate line $c$ & $\overline F_q$ & Flow limit of generic line $q$ \\
$S_{base}$ & Power base of the network model & $\overline\theta$ & Absolute phase-angle limit \\
$P^{DC,raw}_{a,y,s,t}$ & Raw workload demand at data center $a$ & $\overline P^{DC}_{a,y}$ & Workload capacity of data center $a$ \\
$\alpha_a$ & Migratable fraction of workload at data center $a$ & $\eta_a$ & Composite computing-to-grid power conversion factor \\
$\gamma^{data}$ & Data volume per unit of migrated workload & $\overline B_{a,b}$ & Data-rate limit of link $(a,b)$ \\
$\lambda^{BW}$ & Bandwidth scaling factor & $L_{a,b}$ & Latency of communication link $(a,b)$ \\
$\overline L$ & Maximum allowable communication latency & $e^{comm}_{a,b}$ & Communication energy per unit of data \\
\(j\) & Alibaba trace-instance index
& \(m(j)\) & Machine type of instance \(j\) \\

\(g(j)\) & GPU type of instance \(j\)
& \(t_j^{\mathrm{start}}\) & Start time of instance \(j\) \\

\(t_j^{\mathrm{end}}\) & End time of instance \(j\)
& \(b_t\) & Starting time of interval \(t\) \\

$P_j^{\mathrm{CG}}$ & Average CPU--GPU power of instance $j$
& $p_{C,m(j)}^{\mathrm{core}}$ & Dynamic power per vCPU core \\

$P_m^{\mathrm{CPUfull}}$ & Input power of machine type $m$ under full CPU load
& $P_m^{\mathrm{idle}}$ & Idle input power of machine type $m$ \\

$C_m$ & Number of vCPU cores of machine type $m$
& $P_{g(j)}^{\mathrm{rated}}$ & Rated power of GPU type $g(j)$ \\

$L_{j,t}$ & Overlap duration between instance $j$ and interval $t$
& $E_t^{\mathrm{CG}}$ & CPU--GPU energy in interval $t$ \\

$\overline P_t^{\mathrm{CG}}$ & Average CPU--GPU power in interval $t$
& $P_{\mathrm{CG}}^{95}$ & 95th percentile of the average-power series \\

$\widehat u_t$ & Normalized data-center load profile
& $r^L$ & Annual conventional-load growth rate \\

$r^R$ & Annual renewable-growth rate
& $r_a^{DC}$ & Annual workload-growth rate of data center $a$ \\

$r_a^{cap}$ & Annual capacity-growth rate of data center $a$
& $T_c^{build}$ & Construction time of candidate line $c$ \\
$\mathcal M_c$ & Big-M constant of candidate line $c$ & $\kappa$ & Big-M scaling factor \\
$r^{disc}$ & Annual discount rate & $\beta_y$ & Discount factor for year $y$ \\
$I_c$ & Investment cost of candidate line $c$ & $B_y$ & Investment budget in year $y$ \\
$w_s$ & Annual weight of representative day $s$ & $\Delta t$ & Duration of one time interval \\
$c^G_g$ & Variable generation cost of generator $g$ & $e_g$ & Carbon-emission factor of generator $g$ \\
$\pi^{\mathrm{CO}_2}$ & Carbon price & $c^{curt}$ & Renewable-curtailment penalty \\
$c^{shed}$ & Unserved-workload penalty & $c^{work}$ & Base workload-migration cost \\
$c^{data}$ & Data-transfer cost & $\pi^{ele}$ & Electricity price \\
$c^{mig}_{a,b}$ & Unit migration cost on link $(a,b)$ & $C^{inv}$ & Discounted transmission-investment cost \\
$C^{op}$ & Discounted operating cost & $\phi(s)$ & Data-center profile assigned to day $s$ \\
$P_{y,s,t}$ & Generic power term for annual scaling & $E_y$ & Annual energy in year $y$ \\
$\ell^{load}_{q,y,s,t}$ & Absolute loading rate of line $q$ & $L$ & Number of lines built by case C3 \\
\end{longtable}
\endgroup
\setcounter{table}{0}
\renewcommand{\theHtable}{main.\arabic{table}}

\section{Introduction}

Electricity consumption by data centers (DCs) is projected to rise from approximately 485 TWh in 2025 to 950 TWh in 2030, with artificial intelligence (AI) as the main source of growth \citep{IEA2026KeyQuestions}. Transmission expansion, however, proceeds on a much longer timescale. Evidence from China and the United States indicates that major transmission projects can require approximately 1--17 years to complete, depending on project scope and regulatory procedures \citep{ChinaGov2020Zhangbei,NEA2022ChuanYu,DOE2024CITAP,DOE2024Transmission}. Such long lead times may prevent transmission expansion from keeping pace with rapid DC-load growth. A spatial mismatch also exists. DC demand is often concentrated near major load centers, whereas wind and solar resources may be located in remote regions. Transmission constraints can consequently leave DC demand unserved while renewable generation is curtailed elsewhere. Physical expansion remains necessary to provide durable transfer capacity. Given its long lead time, flexible demand-side resources, particularly workload migration, should also be considered to improve the utilization of the existing network during the transition.

Unlike conventional loads, DCs can adjust part of their electricity demand by scheduling delay-tolerant computing tasks. Trace-based assessment of production workloads provides empirical evidence for this flexibility while accounting for task dependencies and service constraints \citep{Cao2022DataDriven}.

Given this load flexibility, a substantial body of research has examined the operational effects of coordinating DCs with power systems. At shorter timescales, DCs have been evaluated as providers of frequency regulation through UPS capacity, server-power control, and coordinated IT-cooling control \citep{Alapera2018Dynamic,Wang2019Frequency,Fu2020Assessments}. Spatiotemporal workload regulation can improve network operating flexibility \citep{Chen2021Operational,Wu2023Incentivizing}, while inter-data-center migration can reduce renewable-energy curtailment and associated carbon emissions \citep{Zheng2020Migration}. Network-constrained studies further show that nodal workload redistribution can relieve congestion, reduce operating costs, and affect locational prices \citep{Wan2025GridOperational,Ye2026Harvesting,Tsiligkaridis2025Distributed}. Zhang et al.~\citep{Zhang2025Mitigating} reported reductions in data-center-induced load fluctuations and line occupancy through coordinated scheduling, while Mansouri et al.~\citep{Mansouri2026Flexibility} used geographically distributed DCs as flexibility providers in congestion management. These studies establish the operational value of workload migration, but they generally take network investment as fixed.

Planning-oriented research remains more limited. Generic demand response has been incorporated into generation and transmission expansion planning \citep{Yang2024Chance,Wei2025Expansion}. For DCs, Chen et al.~\citep{Chen2021Aggregated} aggregated the data network as a virtual power network and incorporated its demand response into generation and transmission expansion planning. However, staged transmission-line construction and commissioning over a multi-year horizon were not represented. Zeng et al.~\citep{Zeng2024BilevelPlanning} and Wan et al.~\citep{Wan2023Transition} examined how data-center flexibility affects investment and low-carbon transition pathways under electricity and carbon policies. Other studies have jointly optimized network investment, data-center or renewable siting, and flexible operation under uncertainty \citep{Dong2024StochasticPlanning,Xiao2026Integrated,Chen2024JointExpansion}. Chen and Zheng~\citep{ChenZheng2026DeferShift} recently evaluated temporal deferral and geographical shifting of AI workloads in a capacity-expansion model, but the investment decisions were primarily associated with additional generation capacity.

Taken together, existing studies have not systematically determined how spatial workload flexibility affects candidate-line selection, construction timing, and staged commissioning when AI-driven DC demand continues to grow, physical lines are subject to construction lead times, and workload migration is constrained by communication bandwidth and latency.

To address this gap, this study co-optimizes workload migration and staged physical transmission expansion over a multi-year horizon. Candidate-line construction starts and commissioning delays are represented together with annual growth in conventional load, renewable availability, and data-center demand. Four cases compare systems with neither resource, VTL only, transmission expansion only, and both resources, thereby identifying their individual and coordinated effects. A counterfactual with VTL enabled under a fixed expansion plan further separates the operating and planning values of VTL. Sensitivity analysis then identifies the conditions under which VTL is valuable and physical reinforcement remains necessary.

This study makes three contributions:

\begin{enumerate}
\item Workload migration among geographically distributed DCs is modeled as a virtual transmission line that redistributes nodal electricity demand under workload-balance, bandwidth, and latency constraints without increasing physical transfer capacity.

\item A multi-year mixed-integer linear programming model is developed to co-optimize VTL operation and staged transmission expansion. The model explicitly represents candidate-line selection, construction timing, commissioning delays, and annual growth in conventional load, renewable availability, and data-center demand.

\item Four planning cases and a fixed-expansion-plan counterfactual are designed to distinguish the individual and coordinated effects of VTL and physical transmission expansion and to separate the operating value of VTL from its planning value. Sensitivity analysis further identifies the conditions under which VTL is valuable and physical reinforcement remains necessary.
\end{enumerate}

\section{Transmission expansion planning model with data-center virtual transmission resources}

\begin{figure}[pos=htbp]
\centering
\fitgraphic{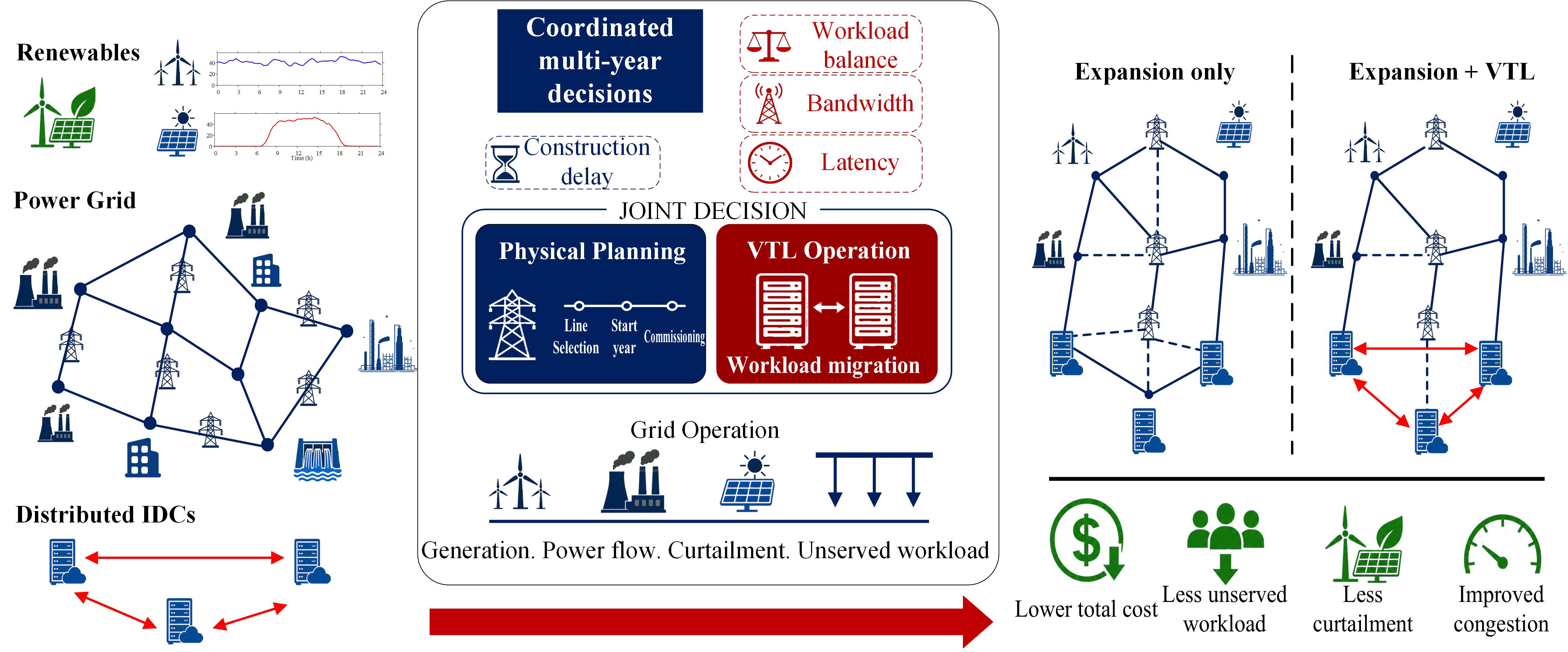}
\caption{Framework for VTL-aware multi-year transmission expansion planning. Workload migration among geographically distributed DCs is coordinated with physical transmission expansion under construction lead-time, workload-balance, bandwidth, and latency constraints.}
\label{fig:framework}
\end{figure}

The modeled system comprises a physical power grid, conventional and renewable generators, conventional loads, DCs, and an inter-DCs communication network. Candidate transmission lines increase the physical transfer capacity of the grid, whereas workload migration redistributes power-equivalent workloads among DCs and thereby changes the spatial distribution of nodal demand. As illustrated in Fig.~\ref{fig:framework}, transmission expansion and VTL operation are coordinated within a multi-year planning framework. Transmission planning determines line selection and construction timing, while VTL operation reallocates workloads among DCs subject to workload-balance, bandwidth, and latency constraints. These decisions jointly affect generation dispatch, power flows, renewable curtailment, and unserved workload. However, only physical transmission lines carry electrical energy; VTLs provide spatial demand flexibility without transmitting electricity.

\subsection{Model assumptions}

To maintain a tractable mixed-integer linear formulation, the following assumptions are adopted.

First, power flows are represented by a DCPF model. Line losses, reactive power, and bus-voltage constraints are ignored, and only the relationship among active power, phase angles, and line flows is modeled. Because this study focuses on transmission expansion planning and line congestion, DCPF captures the principal active-power transfer constraints.

Second, candidate transmission-line construction decisions are represented by binary variables. If a candidate line starts construction in a given year, it becomes available only after a specified construction period. Before commissioning, the line carries no flow; after commissioning, it participates in power-flow allocation and is subject to its capacity limit.

Third, conventional generators are modeled with continuous output variables. Output upper and lower limits, linear generation costs, and carbon-emission factors are considered. Unit commitment variables, minimum up/down times, ramping constraints, and reserve constraints are not included, because the focus of this paper is medium- and long-term transmission expansion rather than short-term unit commitment.

Fourth, renewable generation is represented by available-output profiles. Renewable power in each period can either be consumed by the system or curtailed because of network constraints or insufficient consumption capability. Curtailment is penalized in the objective function.

Fifth, DCs are mapped to the power system through an equivalent aggregated-load model. Computing tasks are represented as power-equivalent workloads. The formulation considers spatial workload migration as a demand-response resource. Joint temporal and spatial flexibility is outside the present scope.

Finally, data-center workloads should be served as much as possible. To avoid infeasibility when data-center load grows rapidly or grid capacity is constrained, an unserved data-center load variable is introduced and assigned a high penalty cost. This variable is used only when the system cannot fully satisfy data-center demand.

\subsection{Sets and decision variables}

All sets, indices, variables, and parameters used below are summarized in the Nomenclature.

\subsection{Constraints}

\subsubsection{Multi-year load, renewable, and data-center demand growth}

The model scales conventional load, renewable availability, DC workload, and executable DC capacity by planning year. The four growth relations are:
\begin{equation}
P^L_{n,y,s,t}=P^L_{n,0,s,t}\left(1+r^L\right)^{y-1}.
\label{eq:load_growth}
\end{equation}

\begin{equation}
\overline{P}^R_{r,y,s,t}=\overline{P}^R_{r,0,s,t}\left(1+r^R\right)^{y-1}.
\label{eq:renewable_growth}
\end{equation}

\begin{equation}
P^{DC,raw}_{a,y,s,t}=P^{DC,raw}_{a,0,s,t}\left(1+r^{DC}_a\right)^{y-1}.
\label{eq:DCs_growth}
\end{equation}

\begin{equation}
\overline{P}^{DC}_{a,y}=\overline{P}^{DC}_{a,0}\left(1+r^{cap}_a\right)^{y-1}.
\label{eq:DCs_capacity_growth}
\end{equation}
\subsubsection{Conventional generator output constraints}

Conventional generators are represented by continuous output variables and satisfy upper and lower output limits:
\begin{equation}
\underline{P}^{G}_{g}\leq P^{G}_{g,y,s,t}\leq \overline{P}^{G}_{g},
\quad \forall g\in \mathcal{G}, y\in\mathcal{Y}, s\in\mathcal{S}, t\in\mathcal{T}.
\label{eq:gen_bounds}
\end{equation}
\subsubsection{Renewable utilization and curtailment constraints}

For each renewable generator, available output is split into consumed output and curtailed output:
\begin{equation}
P^R_{r,y,s,t}+P^{curt}_{r,y,s,t}=\overline{P}^{R}_{r,y,s,t},
\quad \forall r\in\mathcal{R}, y\in\mathcal{Y}, s\in\mathcal{S}, t\in\mathcal{T}.
\label{eq:renewable_balance}
\end{equation}
\subsubsection{DCPF constraints for existing transmission lines}

The DCPF relation for an existing line is:
\begin{equation}
F^E_{e,y,s,t}=b_e\left(\theta_{i(e),y,s,t}-\theta_{j(e),y,s,t}\right),
\quad \forall e\in\mathcal{E},y\in\mathcal{Y},s\in\mathcal{S},t\in\mathcal{T}.
\label{eq:existing_flow}
\end{equation}

It is also constrained by the line capacity:
\begin{equation}
-\overline{F}_e\leq F^E_{e,y,s,t}\leq \overline{F}_e,
\quad \forall e\in\mathcal{E},y\in\mathcal{Y},s\in\mathcal{S},t\in\mathcal{T}.
\label{eq:existing_capacity}
\end{equation}

\subsubsection{Candidate-line construction and service-status constraints}

Candidate-line construction is represented by binary variables. If candidate line $c$ starts construction in year $\tau$ and has a construction duration of $T^{build}_c$, it becomes available only in year $\tau+T^{build}_c$ and thereafter. Thus, the service status of a candidate line in year $y$ is:
\begin{equation}
z_{c,y}=\sum_{\tau\in\mathcal{Y}:\,\tau+T^{build}_c\leq y} x_{c,\tau},
\quad \forall c\in\mathcal{C}, y\in\mathcal{Y}.
\label{eq:service_status}
\end{equation}
If no previous construction decision satisfies the lead-time requirement, the line is unavailable in year $y$:
\begin{equation}
z_{c,y}=0, \quad \text{if } \{\tau\in\mathcal{Y}:\tau+T^{build}_c\leq y\}=\emptyset.
\label{eq:service_unavailable}
\end{equation}

Each candidate line can be built at most once over the planning horizon:
\begin{equation}
\sum_{y\in\mathcal{Y}}x_{c,y}\leq 1,
\quad \forall c\in\mathcal{C}.
\label{eq:build_once}
\end{equation}
In cases where transmission expansion is not allowed, the model further fixes:
\begin{equation}
x_{c,y}=0,\quad z_{c,y}=0,
\quad \forall c\in\mathcal{C}, y\in\mathcal{Y}.
\label{eq:no_build}
\end{equation}

\subsubsection{Candidate-line flow constraints}

Candidate lines can carry flow only after they are commissioned. Therefore:
\begin{equation}
-\overline{F}_c z_{c,y}\leq F^C_{c,y,s,t}\leq \overline{F}_c z_{c,y},
\quad \forall c\in\mathcal{C},y\in\mathcal{Y},s\in\mathcal{S},t\in\mathcal{T}.
\label{eq:candidate_capacity}
\end{equation}

When a candidate line is in service, its flow satisfies the DCPF equation. When it is not in service, the capacity constraint fixes its flow to zero. To preserve linearity, a Big-M formulation links candidate-line flow and service status:
\begin{align}
F^C_{c,y,s,t}-b_c\left(\theta_{i(c),y,s,t}-\theta_{j(c),y,s,t}\right)
&\leq \mathcal{M}_c\left(1-z_{c,y}\right), \label{eq:candidate_big_m_upper}\\
-F^C_{c,y,s,t}+b_c\left(\theta_{i(c),y,s,t}-\theta_{j(c),y,s,t}\right)
&\leq \mathcal{M}_c\left(1-z_{c,y}\right). \label{eq:candidate_big_m_lower}
\end{align}
The candidate-line coefficient is computed as $b_c=S_{base}/X_c$. The Big-M constant is:
\begin{equation}
\mathcal{M}_c=\kappa\left(2|b_c|\overline{\theta}+\overline{F}_c\right),
\label{eq:big_m}
\end{equation}
\subsubsection{Reference-bus phase-angle constraints}

To ensure a unique solution of the DCPF equations, one reference bus $n_0$ is selected and its phase angle is fixed to zero:
\begin{equation}
\theta_{n_0,y,s,t}=0,\quad \forall y\in\mathcal{Y},s\in\mathcal{S},t\in\mathcal{T}.
\label{eq:slack_bus}
\end{equation}
All bus phase angles also satisfy:
\begin{equation}
-\overline{\theta}\leq \theta_{n,y,s,t}\leq \overline{\theta},
\quad \forall n\in\mathcal{N},y\in\mathcal{Y},s\in\mathcal{S},t\in\mathcal{T}.
\label{eq:theta_bounds}
\end{equation}

\subsubsection{Data-center workload migration and executed-load constraints}

Data-center workload migration is described through directed communication links. For DC $a$, the workload finally executed locally equals raw demand minus unserved load and outbound migration, plus inbound migration from other DCs:
\begin{equation}
P^{exec}_{a,y,s,t}
=P^{DC,raw}_{a,y,s,t}
-P^{shed}_{a,y,s,t}
-\sum_{b:(a,b)\in\mathcal{K}}M_{a,b,y,s,t}
+\sum_{b:(b,a)\in\mathcal{K}}M_{b,a,y,s,t}.
\label{eq:DCs_exec}
\end{equation}
The executed workload must be nonnegative and cannot exceed the available data-center capacity:
\begin{equation}
0\leq P^{exec}_{a,y,s,t}\leq \overline{P}^{DC}_{a,y},
\quad \forall a\in\mathcal{D}.
\label{eq:DCs_exec_bounds}
\end{equation}
Unserved workload cannot exceed raw workload demand:
\begin{equation}
0\leq P^{shed}_{a,y,s,t}\leq P^{DC,raw}_{a,y,s,t},
\quad \forall a\in\mathcal{D}.
\label{eq:DCs_shed_bound}
\end{equation}

\subsubsection{Data-center migration-flexibility constraints}

The outbound migrated workload is limited by the migratable fraction:
\begin{equation}
\sum_{b:(a,b)\in\mathcal{K}}M_{a,b,y,s,t}
\leq \alpha_a P^{DC,raw}_{a,y,s,t},
\quad \forall a\in\mathcal{D}.
\label{eq:migration_flex}
\end{equation}
In no-VTL cases, the right-hand side is set to zero:
\begin{equation}
\sum_{b:(a,b)\in\mathcal{K}}M_{a,b,y,s,t}=0,
\quad \forall a\in\mathcal{D},
\label{eq:no_vtl}
\end{equation}
which means that workload migration among DCs is not allowed.

\subsubsection{Communication bandwidth and latency constraints}

Communication capacity limits the migration rate:
\begin{equation}
\gamma^{data}M_{a,b,y,s,t}\leq \lambda^{BW}\overline{B}_{a,b},
\quad \forall (a,b)\in\mathcal{K},y\in\mathcal{Y},s\in\mathcal{S},t\in\mathcal{T}.
\label{eq:bandwidth}
\end{equation}
Links that exceed the latency threshold are disabled:
\begin{equation}
M_{a,b,y,s,t}=0,\quad \text{if } L_{a,b}>\overline{L}.
\label{eq:latency}
\end{equation}

In the current implementation, communication energy is monetized in the migration cost and reported as an output statistic. It is not assigned as an additional electric load at a grid node.

\subsubsection{Nodal power-balance constraints}

For each grid bus, conventional generation and renewable generation must equal conventional load, data-center electric load, and net outgoing line flow. The nodal power-balance constraint is:
\begin{align}
&\sum_{g\in\mathcal{G}_n}P^G_{g,y,s,t}
+\sum_{r\in\mathcal{R}_n}P^R_{r,y,s,t}
-P^L_{n,y,s,t}
-\sum_{a\in\mathcal{D}_n}\eta_a P^{exec}_{a,y,s,t} \notag\\
&=
\sum_{e:i(e)=n}F^E_{e,y,s,t}
-\sum_{e:j(e)=n}F^E_{e,y,s,t}
+\sum_{c:i(c)=n}F^C_{c,y,s,t}
-\sum_{c:j(c)=n}F^C_{c,y,s,t}.
\label{eq:power_balance}
\end{align}
The right-hand side is the net outgoing flow on existing and commissioned candidate lines.

This constraint is the core coupling between the power grid and data-center migration behavior. A VTL does not directly change the physical grid topology and does not add a physical transmission line. Instead, it changes the data-center executed load at different nodes, thereby reshaping nodal net injections and physical line-flow distributions.

\subsubsection{Annual investment budget constraints}

If an annual transmission investment budget is considered, candidate-line investment in year $y$ cannot exceed the available budget:
\begin{equation}
\sum_{c\in\mathcal{C}}I_c x_{c,y}\leq B_y,
\quad \forall y\in\mathcal{Y}.
\label{eq:annual_budget}
\end{equation}
If no annual budget is specified, this constraint is disabled.

\subsection{Objective function and solution method}

The objective is to minimize total discounted cost over the multi-year planning horizon. The objective consists of two parts: transmission-line investment cost and representative-day operating cost. The discount factor is defined as:
\begin{equation}
\beta_y=\frac{1}{(1+r^{disc})^{y-1}},
\label{eq:discount_factor}
\end{equation}
The candidate-line investment cost is:
\begin{equation}
C^{inv}=\sum_{y\in\mathcal{Y}}\beta_y\sum_{c\in\mathcal{C}}I_c x_{c,y}.
\label{eq:investment_cost}
\end{equation}

Operating cost includes conventional generation, carbon emissions, renewable curtailment, unserved DC load, workload migration, and communication. It is:
\begin{align}
C^{op}=&\sum_{y\in\mathcal{Y}}\beta_y
\sum_{s\in\mathcal{S}}w_s
\sum_{t\in\mathcal{T}}\Delta t
\Bigg[
\sum_{g\in\mathcal{G}}\left(c^G_g+\pi^{\mathrm{CO}_2}e_g\right)P^G_{g,y,s,t} \notag\\
&+\sum_{r\in\mathcal{R}}c^{curt}P^{curt}_{r,y,s,t}
+\sum_{a\in\mathcal{D}}c^{shed}P^{shed}_{a,y,s,t}
+\sum_{(a,b)\in\mathcal{K}}c^{mig}_{a,b}M_{a,b,y,s,t}
\Bigg].
\label{eq:operation_cost}
\end{align}
The unit migration cost consists of the base workload migration cost, communication energy cost, and data traffic cost:
\begin{equation}
c^{mig}_{a,b}
=c^{work}
+\gamma^{data} e^{comm}_{a,b}\pi^{ele}
+\gamma^{data}c^{data},
\label{eq:migration_cost}
\end{equation}
The complete objective function is:
\begin{equation}
\min \; C^{inv}+C^{op}.
\label{eq:objective}
\end{equation}
This objective jointly optimizes physical transmission-line construction and data-center workload migration under grid operation constraints, transmission construction constraints, data-center service-quality constraints, and communication-network constraints, so that total discounted multi-year cost is minimized.

\section{Case study}

\subsection{Test system}

\begin{figure}[pos=htbp]
\centering
\fitgraphic{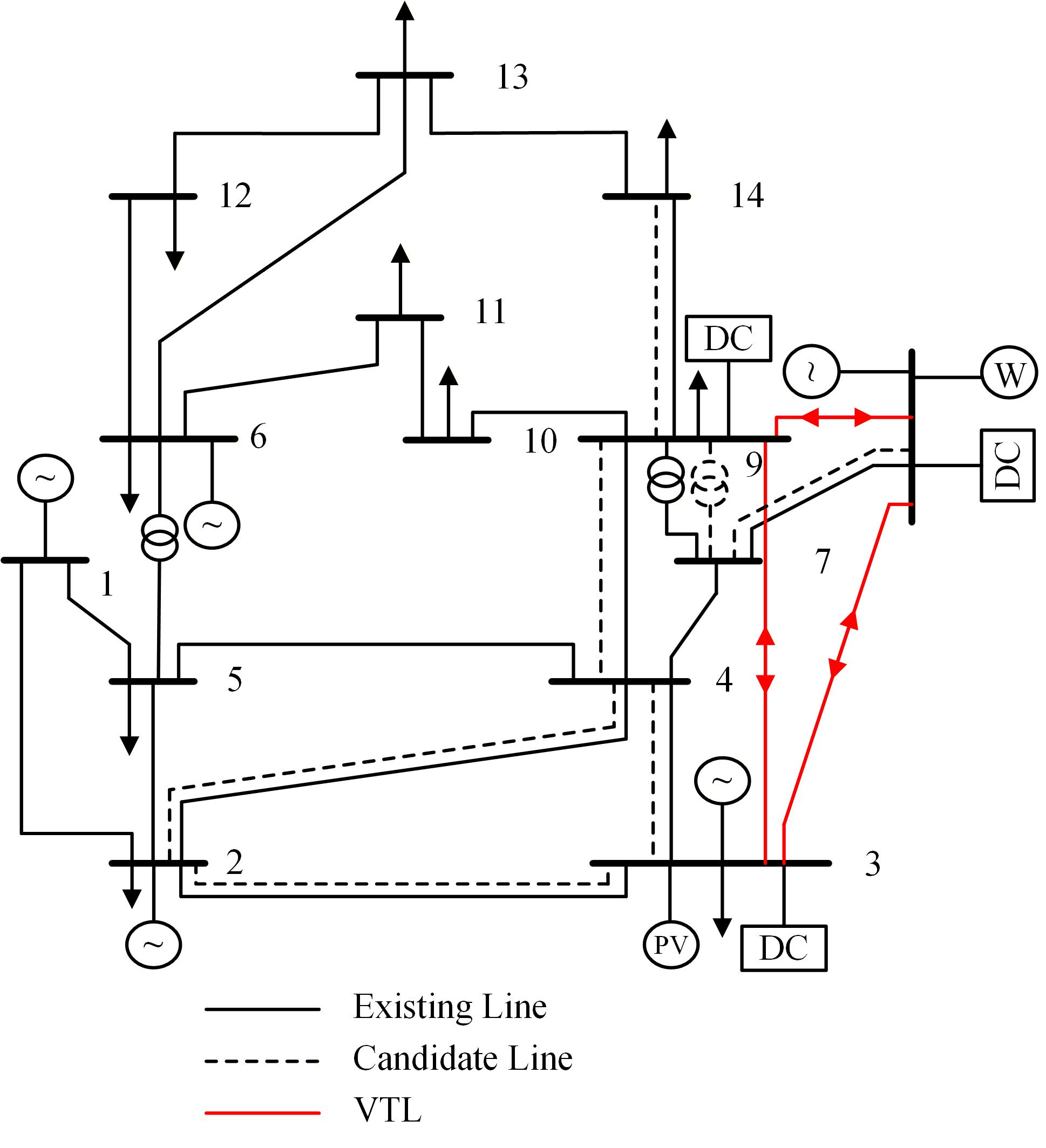}
\caption{Modified IEEE 14-bus test system showing renewable generators, DCs, existing lines, candidate lines, and VTL links.}
\label{fig:test_system}
\end{figure}

The test system is a modified IEEE 14-bus network with candidate lines, renewable generators, and three DCs. Wind generation and DC2 are connected at bus 8, photovoltaic generation and DC3 at bus 3, and DC1 at the load-center bus 9. This placement creates renewable-rich sending areas and a load-center receiving area. Existing lines, candidate lines, and VTL links are distinguished in Fig.~\ref{fig:test_system}.

Each candidate line has specified terminal buses, reactance, capacity, investment cost, and a two-year construction lead time. A line carries no power until its lead time has elapsed. Appendix Tables~\ref{tab:existing_lines}--\ref{tab:comm_links} report the complete existing-line, candidate-line, generator, bus-load, DC, and communication-link data used by the model.

\subsection{Data}

Wind, solar, and conventional-load profiles are taken from the \textit{Renewable Energy and Electricity Demand Time Series Dataset with Exogenous Variables at 5-minute Interval} \citep{RojasOrtega2023Dataset}. Data from January 1, 2019 through December 31, 2021 are resampled to 15-minute intervals and grouped by season. Within each season, high-renewable days are selected using the upper tertile of daily wind and solar output. High-load days are then selected from the remaining days using daily peak demand, and the rest form the normal class. For each class, normalized 96-point photovoltaic, wind, and load profiles are combined into a feature vector; the observed day closest to the class mean in Euclidean distance is retained. The procedure yields 12 representative days. Their weights, reported in Appendix Table~\ref{tab:representative_weights}, sum to 365 days/year.

Direct facility-level power measurements are unavailable in the dataset. We therefore construct a 15-minute computing-power profile from the \textit{Alibaba 2020 GPU-cluster} trace, which records training and inference workloads from a production cluster \citep{Weng2022MLaaS}. Instance-operation and resource-usage records are linked through \texttt{worker\_name} to obtain the start time, end time, lifecycle-average CPU usage, lifecycle-average GPU usage, and GPU type of each instance.

The average CPU-GPU computing power of instance \(j\) is estimated as
\begin{equation}
P_j^{\mathrm{CG}}
=
p_{C,m(j)}^{\mathrm{core}}
\frac{\mathrm{cpu\_usage}_j}{100}
+
P_{g(j)}^{\mathrm{rated}}
\frac{\mathrm{gpu\_wrk\_util}_j}{100},
\label{eq:instance_cg_power}
\end{equation}
where \(\mathrm{cpu\_usage}_j/100\) is the average number of vCPU cores used by the instance, and \(\mathrm{gpu\_wrk\_util}_j/100\) is the average equivalent number of GPUs used. The dynamic power per vCPU core, \(p_{C,m(j)}^{\mathrm{core}}\), is obtained by dividing the difference between the full-load and idle power of machine type \(m(j)\) by its total number of vCPU cores. The rated power \(P_{g(j)}^{\mathrm{rated}}\) is matched to the recorded GPU type using the corresponding manufacturer specification.

The trace provides lifecycle-average resource usage rather than an instantaneous power series. Hence, \(P_j^{\mathrm{CG}}\) is treated as the average computing power over the lifetime of instance \(j\). The observation period is divided into consecutive 15-minute intervals. The overlap duration between instance \(j\) and interval \(t\) is
\begin{equation}
L_{j,t}
=
\max
\left\{
0,\,
\min
\left(
t_j^{\mathrm{end}}, b_t+900
\right)
-
\max
\left(
t_j^{\mathrm{start}}, b_t
\right)
\right\},
\label{eq:interval_overlap}
\end{equation}
where \(b_t\) is the starting time of interval \(t\), and \(L_{j,t}\) is measured in seconds. This overlap-based allocation assigns the energy of an instance to every interval in which it operates, rather than assigning the entire record to the interval containing its midpoint.

The computing energy and average computing power in interval \(t\) are calculated as
\begin{equation}
E_t^{\mathrm{CG}}
=
\sum_j
\frac{P_j^{\mathrm{CG}}}{1000}
\frac{L_{j,t}}{3600},
\qquad
\overline P_t^{\mathrm{CG}}
=
\frac{E_t^{\mathrm{CG}}}{0.25},
\label{eq:interval_cg_energy}
\end{equation}
where \(E_t^{\mathrm{CG}}\) is measured in kWh and \(\overline P_t^{\mathrm{CG}}\) in kW. Arranging \(\overline P_t^{\mathrm{CG}}\) chronologically yields the 15-minute computing-power profile. When the intervals cover the complete lifetime of every instance, the interval-level energy total equals the sum of the corresponding instance-level energy estimates.

To obtain the dimensionless load shape required by the planning model, the average-power series is normalized using its 95th-percentile value over the complete observation period:
\begin{equation}
\widehat u_t
=
\min
\left(
1,\,
\frac{\overline P_t^{\mathrm{CG}}}
{P_{\mathrm{CG}}^{95}}
\right),
\label{eq:normalized_dc_profile}
\end{equation}
where \(P_{\mathrm{CG}}^{95}\) denotes the 95th percentile of the full 15-minute average-power series. Using a common reference value preserves the relative differences among days while limiting the influence of extreme observations.

Complete days are ranked according to their daily computing energy and divided into three equal-sized groups using empirical tertiles, corresponding to low-, normal-, and peak-load days. For each group, the 96-point normalized profiles are averaged to obtain a centroid, and the observed day with the shortest Euclidean distance to that centroid is retained as the representative day. Normal profiles are assigned to the normal and high-renewable power-system scenarios, whereas the peak profile is assigned to the peak-load scenario. The resulting normalized profile is mapped to the installed capacity of each DC as
\begin{equation}
P^{\mathrm{DC,raw}}_{a,0,s,t}
=
\overline P^{\mathrm{DC}}_{a,0}
\widehat u_{\phi(s),t},
\label{eq:DCs_capacity_mapping}
\end{equation}
where \(\overline P^{\mathrm{DC}}_{a,0}\) is the initial capacity of DC \(a\), and \(\phi(s)\) maps each power-system scenario to its corresponding data-center load profile.

The trace-derived series represents CPU--GPU computing power rather than the total electricity demand of a DC. In the nodal power-balance constraint, grid-side DC demand is obtained by multiplying the executed compute-side workload by \(\eta_a\), as shown in Eq.~\eqref{eq:power_balance}. The coefficient \(\eta_a\) is a composite computing-to-grid conversion factor that accounts for auxiliary IT equipment, including memory, storage, networking equipment, and server fans, as well as facility overheads associated with cooling and power distribution.

\subsection{Case settings}

To distinguish physical transmission construction from virtual transmission lines, four core comparison cases are defined:

\begin{itemize}
\item \textbf{Case C0: No-build and no-VTL.} This case allows neither candidate-line construction nor workload migration among DCs. It characterizes system operating stress without either resource.
\item \textbf{Case C1: VTL-only.} This case does not allow new candidate transmission lines, but allows workload migration among DCs. It is used to evaluate the operational improvement provided by data-center virtual transmission resources when no new physical line is added.
\item \textbf{Case C2: Build-only.} This case allows candidate-line construction but not workload migration. It is the conventional transmission expansion planning benchmark.
\item \textbf{Case C3: VTL-aware build.} This case allows both candidate-line construction and data-center workload migration. It is used to evaluate the system value of coordinated optimization between physical transmission expansion and data-center virtual transmission resources.
\end{itemize}

Case C2-fixed+VTL is a counterfactual used only for mechanism identification. It fixes the line construction decisions from C2 and then enables VTL operation. The scalar parameters used directly in the objective function and model-wide constraints are summarized in Table~\ref{tab:parameters}.

\begin{table}[pos=htbp]
\caption{Core optimization parameters.}
\label{tab:parameters}
\centering
\setlength{\tabcolsep}{4pt}
\begin{tabular}{@{}llr@{\hspace{10pt}}llr@{}}
\toprule
Parameter & Symbol & Value & Parameter & Symbol & Value \\
\midrule
Planning horizon
& $|\mathcal{Y}|$
& 8 years
& Power base
& $S_{\mathrm{base}}$
& 100 MVA \\

Discount rate
& $r^{\mathrm{disc}}$
& 0.05/year
& Carbon price
& $\pi^{\mathrm{CO}_2}$
& 100 yuan/tCO$_2$ \\

Curtailment penalty
& $c^{\mathrm{curt}}$
& 1000 yuan/MWh
& Unserved DC penalty
& $c^{\mathrm{shed}}$
& 20,000 yuan/MWh \\

Electricity price
& $\pi^{\mathrm{ele}}$
& 0.7 yuan/kWh
& Data volume
& $\gamma^{\mathrm{data}}$
& 50 GB/MWh \\

Workload migration cost
& $c^{\mathrm{work}}$
& 0 yuan/MWh
& Data-transfer cost
& $c^{\mathrm{data}}$
& 0 yuan/GB \\

Phase-angle bound
& $\overline{\theta}$
& 3.14 rad
& Big-M scale
& $\kappa$
& 1.2 \\

Annual investment budget
& $B_y$
& 250 million yuan/year
& &
& \\

\bottomrule
\end{tabular}
\end{table}

The high unserved-load penalty discourages the optimizer from using load loss as a routine operating option. Because its monetary value can mechanically affect reported cost savings, the sensitivity analysis also reports the physical reduction in unserved DC load. Table~\ref{tab:sensitivity_ranges} gives all tested ranges.

To examine whether the formulation can be applied beyond the modified IEEE 14-bus system, the same four cases are also implemented on the IEEE 118-bus network from PGLib-OPF \citep{Babaeinejadsarookolaee2019PGLib}. This supplementary validation uses a five-year planning horizon, 12 representative days, 19 conventional generators, six renewable plants, six geographically distributed DCs, and 20 candidate transmission corridors. The load, renewable, and DC temporal profiles follow the data-processing procedure described above and are rescaled to the larger system. Candidate corridors are selected before optimization by screening base-network DC power flows and ranking existing non-transformer corridors according to their maximum loading, 95th-percentile loading, and frequency of loading above 90\%. The C0--C3 comparison is then repeated with the same planning and operating constraints to assess the formulation on a larger network and candidate set.

\section{Results and discussion}

The MILP is implemented in Python and solved with Gurobi 13.0.2 on an Intel Xeon Platinum processor with 16 physical cores and 32 logical processors. Each run uses eight solver threads, a relative MILP-gap tolerance of 0.005, and a 14,400-s time limit. The stopping rule is the first occurrence of the prescribed gap or the time limit. C2 and C3 terminate within the gap tolerance after 1813.68 and 2094.14 s, respectively, with final gaps of 0.2576\% and 0.4306\%.

\subsection{Overall planning performance}

Table~\ref{tab:main_results} reports the eight-year planning outcomes and the fixed-plan counterfactual.

\begin{table}[pos=htbp]
\caption{Eight-year summary results for the planning cases and the fixed-plan counterfactual.}
\label{tab:main_results}
\centering
\setlength{\tabcolsep}{3.5pt}
\begin{tabular}{@{}lrrrrrrr@{}}
\toprule
Case & Total & Disc. inv. & Lines & Undisc. inv. & Unserved DC & Curtail. & VTL mig. \\
 & (100M yuan) & (100M yuan) &  & (100M yuan) & (GWh) & (GWh) & (GWh) \\
\midrule
C0 & 325.14 & 0.00 & 0 & 0.00 & 1429.90 & 354.27 & 0.00 \\
C1 & 171.79 & 0.00 & 0 & 0.00 & 454.46 & 6.69 & 2084.50 \\
C2 & 138.50 & 7.06 & 4 & 7.60 & 168.80 & 56.37 & 0.00 \\
C3 & 119.87 & 5.53 & 3 & 5.80 & 80.42 & 1.86 & 1980.24 \\
C2-fixed+VTL & 121.41 & 7.06 & 4 & 7.60 & 78.12 & 1.59 & 1967.21 \\
\bottomrule
\end{tabular}
\end{table}

Relative to C2, C3 lowers total discounted cost from 13.850 to 11.987 billion yuan, an approximately 13.45\% reduction based on the rounded table values. Unserved DC load falls from 168.80 to 80.42 GWh (approximately 52.4\%), and renewable curtailment falls from 56.37 to 1.86 GWh (approximately 96.7\%). C3 builds three lines rather than four and reduces discounted investment from 0.706 to 0.553 billion yuan. These changes arise because migration reallocates flexible DC demand toward nodes that can serve it with less curtailment and less stress on constrained corridors. The VTL thus changes the value of candidate lines; it does not add physical transfer capacity.

The optimized line-construction schedules are shown in Fig.~\ref{fig:construction_schedule}; their numerical values are listed in Appendix Table~\ref{tab:construction_schedule}.

\begin{figure}[pos=htbp]
\centering
\fitgraphic{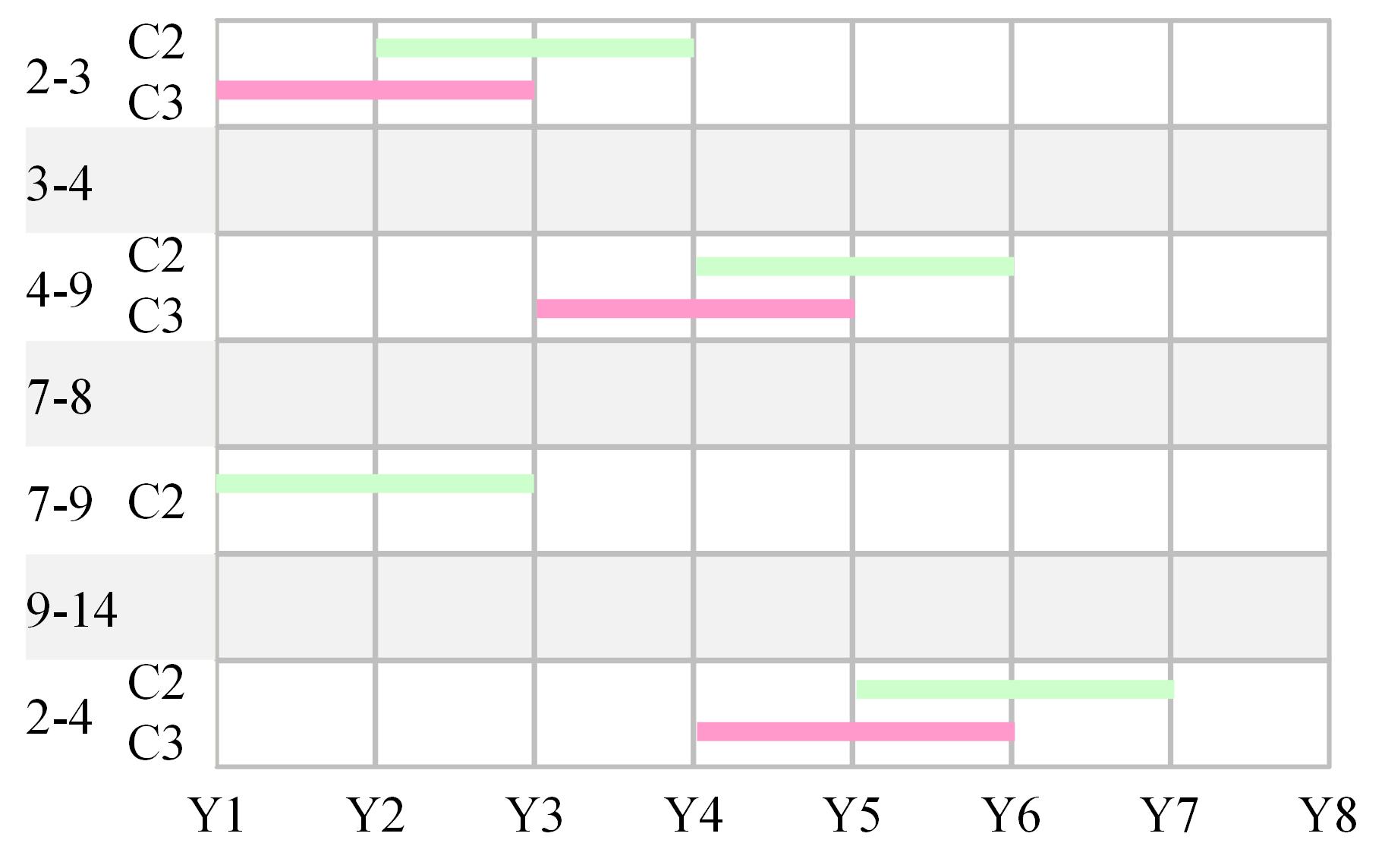}
\caption{Optimized candidate-line construction schedules under the build-only (C2) and VTL-aware build (C3) cases. Each bar extends from the construction start year to the in-service year.}
\label{fig:construction_schedule}
\end{figure}

For the three corridors common to C2 and C3, the construction windows shift from Y2--Y4 to Y1--Y3 for line 2--3, from Y3--Y5 to Y2--Y4 for line 4--9, and from Y4--Y6 to Y3--Y5 for line 2--4. C2 additionally builds line 7--9 over Y1--Y3. In this test system, VTL-aware planning therefore omits one corridor while advancing the three common projects by one year. 

The C1 result sets the operating boundary. VTL alone cuts curtailment and unserved load relative to C0, but its 17.179-billion-yuan cost remains well above C3 and it leaves 454.46 GWh of DC load unserved. Workload migration can relieve local transfer constraints and improve expansion decisions, but continued DC-load growth eventually requires physical reinforcement. The two resources are complementary rather than interchangeable.

\subsection{Annual trajectories}

\begin{figure}[pos=htbp]
\centering
\fitgraphic{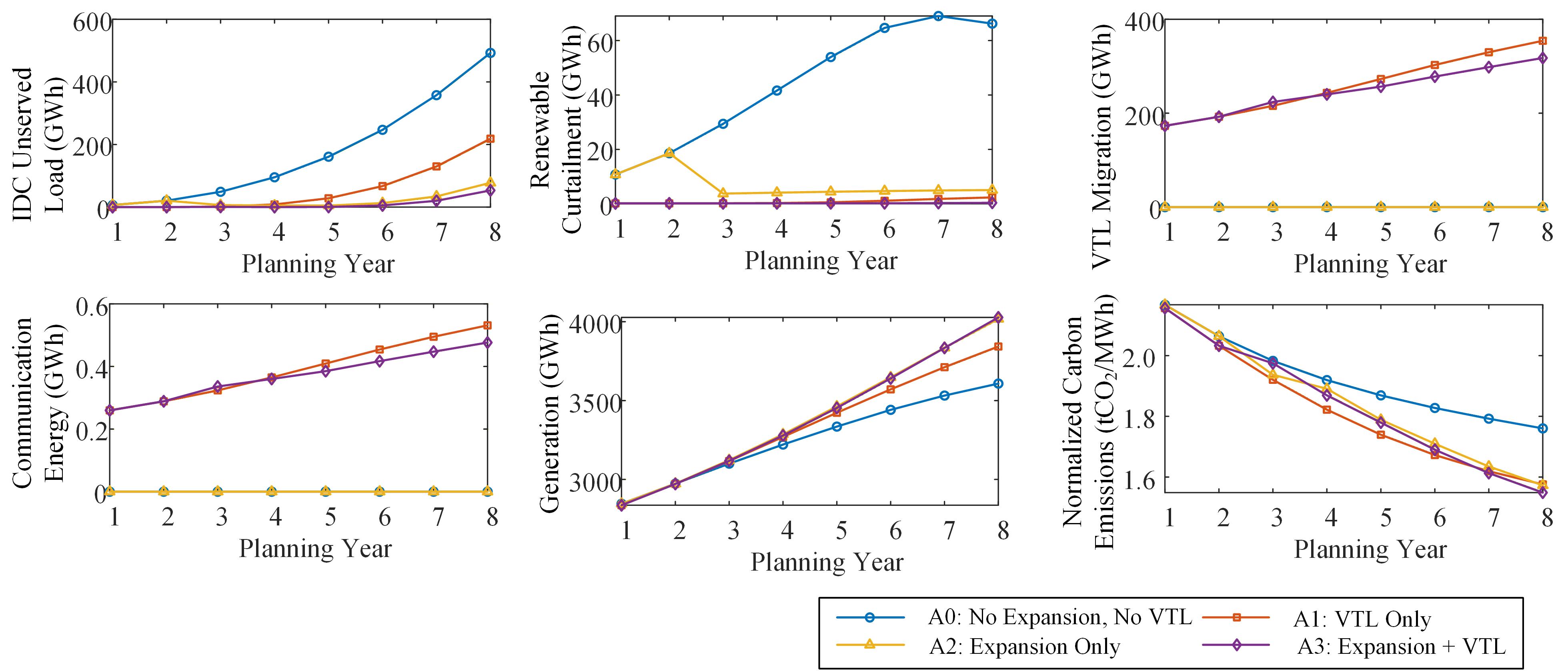}
\caption{Annual operating performance of the four planning cases over the eight-year horizon: unserved DC load, renewable curtailment, migrated workload, communication energy consumption, total generation, and carbon emissions normalized by served DC workload.}
\label{fig:annual_results}
\end{figure}

Figure~\ref{fig:annual_results} compares the annual operating trajectories of the four planning cases.

C1 serves early-year demand with migration, but its unserved DC load rises rapidly in later years as the no-build network reaches its capacity boundary. C3 maintains much lower curtailment and unserved load by coordinating migration with commissioned lines. Migrated workload rises with DC demand in C1 and C3, whereas communication energy remains below 0.6 GWh/year in the plotted runs and is small relative to annual generation.

Total generation increases as more DC load is served. At the same time, carbon emissions normalized by served DC workload decline across the horizon and are lowest for C3 in the later years. The intensity curve indicates that each MWh of served DC workload is associated with less generation-related carbon in the reported normalization.

\subsection{Congestion mechanism}

Line congestion is described using the line loading rate:
\begin{equation}
\ell^{load}_{q,y,s,t}
=\frac{|F_{q,y,s,t}|}{\overline{F}_q}\times 100\%,
\quad q\in\mathcal{E}\cup\mathcal{C},
\label{eq:loading_percent}
\end{equation}
Candidate-line records are included only after commissioning.
Loading records are counted over all lines, years, representative days, and intervals. The raw C0--C3 counts are reported only in Appendix Table~\ref{tab:congestion_counts}, because those cases serve different amounts of load and build different lines. A lower count can therefore reflect more load shedding rather than better congestion management. Likewise, C2 and C3 cannot isolate the operating effect of migration because C3 builds one fewer line.

The C2-fixed+VTL counterfactual holds the physical construction plan constant and changes only VTL availability. Table~\ref{tab:fixed_vtl_congestion} also reports C3 to distinguish this operating comparison from the joint planning outcome.

\begin{table}[pos=htbp]
\caption{Congestion comparison for C2, the fixed-plan VTL counterfactual, and C3.}
\label{tab:fixed_vtl_congestion}
\centering
\begin{tabular}{@{}lrrrrr@{}}
\toprule
Case & Avg. loading (\%) & Max loading (\%) & $\geq$95\% & $\geq$99\% & =100\% \\
\midrule
C2 & 46.73 & 100.00 & 37,737 & 31,712 & 30,299 \\
C2-fixed+VTL & 43.50 & 100.00 & 37,494 & 30,284 & 28,716 \\
C3 & 44.41 & 100.00 & 38,139 & 31,640 & 30,340 \\
\bottomrule
\end{tabular}
\end{table}

With the C2 plan fixed, VTL reduces average loading from 46.73\% to 43.50\%. Records at or above 95\% fall from 37,737 to 37,494, records at or above 99\% from 31,712 to 30,284, and fully loaded records from 30,299 to 28,716. These paired changes identify the operating effect of nodal load redistribution. C3 instead combines that effect with a different investment plan; its congestion counts should not be interpreted as a direct measure of VTL performance.

\subsection{Sensitivity analysis}

\begin{figure}[pos=htbp]
\centering
\fitgraphic{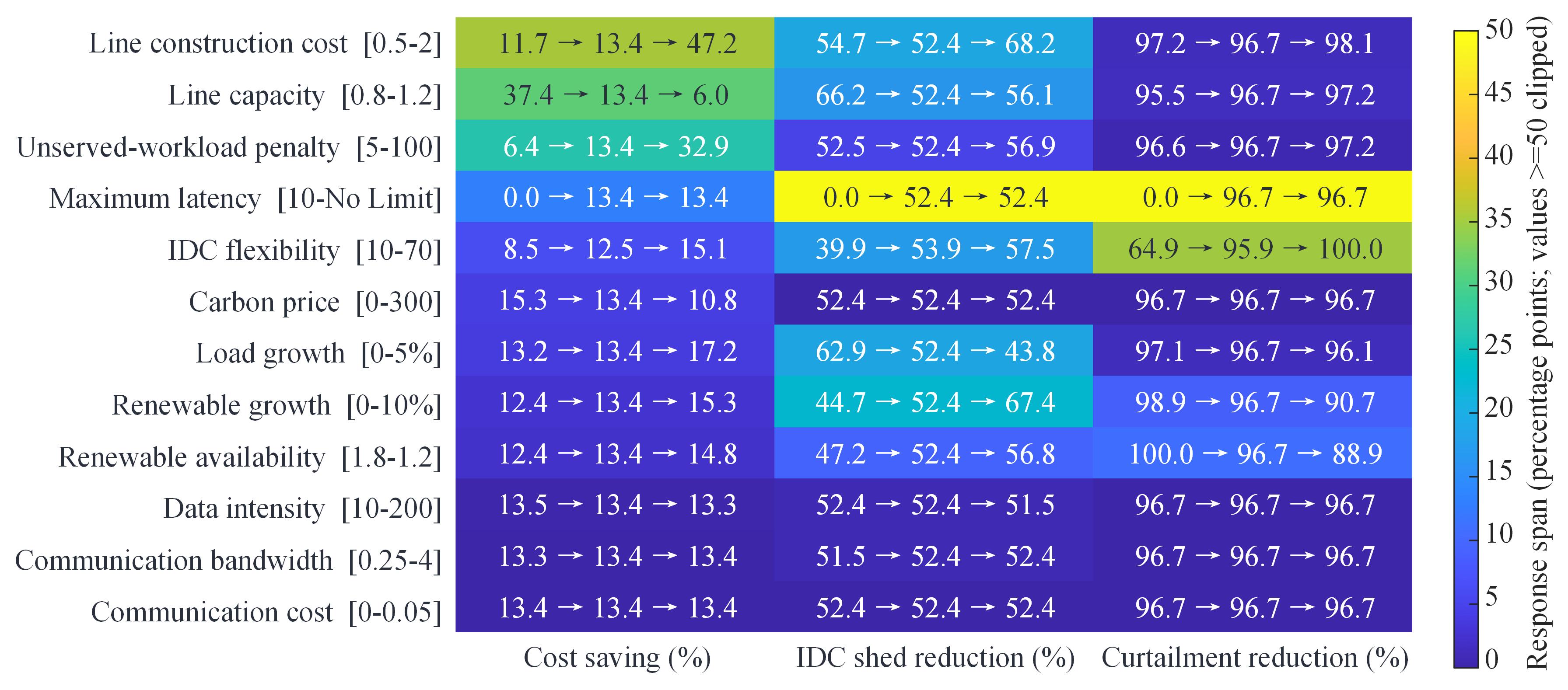}
\caption{Sensitivity of VTL benefits to grid, renewable, DC-load, communication-network, and economic parameters. Each cell lists the response at the lowest tested setting, the baseline, and the highest tested setting (left to right). Color represents the response span in percentage points; spans at or above 50 percentage points are clipped at the upper end of the color scale. The three columns report C3 cost savings, reduction in unserved DC load, and reduction in renewable curtailment relative to C2. The arrows connect tested settings and do not imply linear interpolation.}
\label{fig:sensitivity_heatmap}
\end{figure}

Sensitivity is measured by the C3 cost saving relative to C2, the reduction in unserved DC load, and the reduction in renewable curtailment. Each metric is expressed as a percentage of the corresponding C2 value. Figure~\ref{fig:sensitivity_heatmap} gives the overall response; complete tested settings are moved to Appendix Table~\ref{tab:sensitivity_ranges}.

Across most tested settings, C3 retains positive cost and physical-load benefits, but the magnitude and mechanism differ. Tightening line capacity from 1.2 to 0.8 times the baseline raises the absolute discounted cost saving from 0.695 to 9.287 billion yuan (Appendix Table~\ref{tab:grid_sensitivity}). When transfer capacity is scarce, relocating flexible demand avoids more unserved load and makes better use of renewable output. Raising candidate-line cost likewise increases the value of migration. At twice the baseline cost, however, C3 builds no line and leaves 454.46 GWh unserved. High construction cost therefore increases the apparent substitution value of VTL without making it a complete replacement for the physical grid.

Renewable growth strengthens the curtailment mechanism. As the annual renewable growth rate rises from 0 to 10\%, the absolute curtailment reduction attributable to C3 increases from 28.28 to 151.86 GWh, even though the percentage reduction shown in Fig.~\ref{fig:sensitivity_heatmap} declines at the highest setting because the C2 denominator also changes. The renewable-availability sweep is reported in absolute terms in Appendix Table~\ref{tab:renewable_sensitivity}. At availability scales of 0.8, 1.0, and 1.2, C2--C3 discounted cost savings are 1.788, 1.862, and 1.996 billion yuan, respectively. The corresponding curtailment reductions are 9.49, 54.52, and 155.45 GWh.

The carbon-price sweep uses 0, 50, 100, 200, and 300 yuan/tCO$_2$, with complete results in Appendix Table~\ref{tab:carbon_sensitivity}. The optimized line counts, unserved DC load, and renewable curtailment are unchanged across these five points. The absolute C2--C3 cost saving decreases slightly from 1.864 to 1.860 billion yuan because the carbon-price term changes the monetized operating cost without changing the selected plans in these runs.

Conventional-load growth increases system stress and raises the absolute cost saving from 1.661 to 2.546 billion yuan between the 0 and 4\%/year settings. Greater DC flexibility improves all three response metrics, but the detailed curve flattens at high flexibility as line capacity and the candidate set become the binding constraints. Increasing the unserved-load penalty raises monetized cost savings partly by construction: the same avoided MWh receives a larger price. The physical reduction in unserved DC load must therefore be reported alongside cost.

\begin{figure}[pos=htbp]
\centering
\fitgraphic{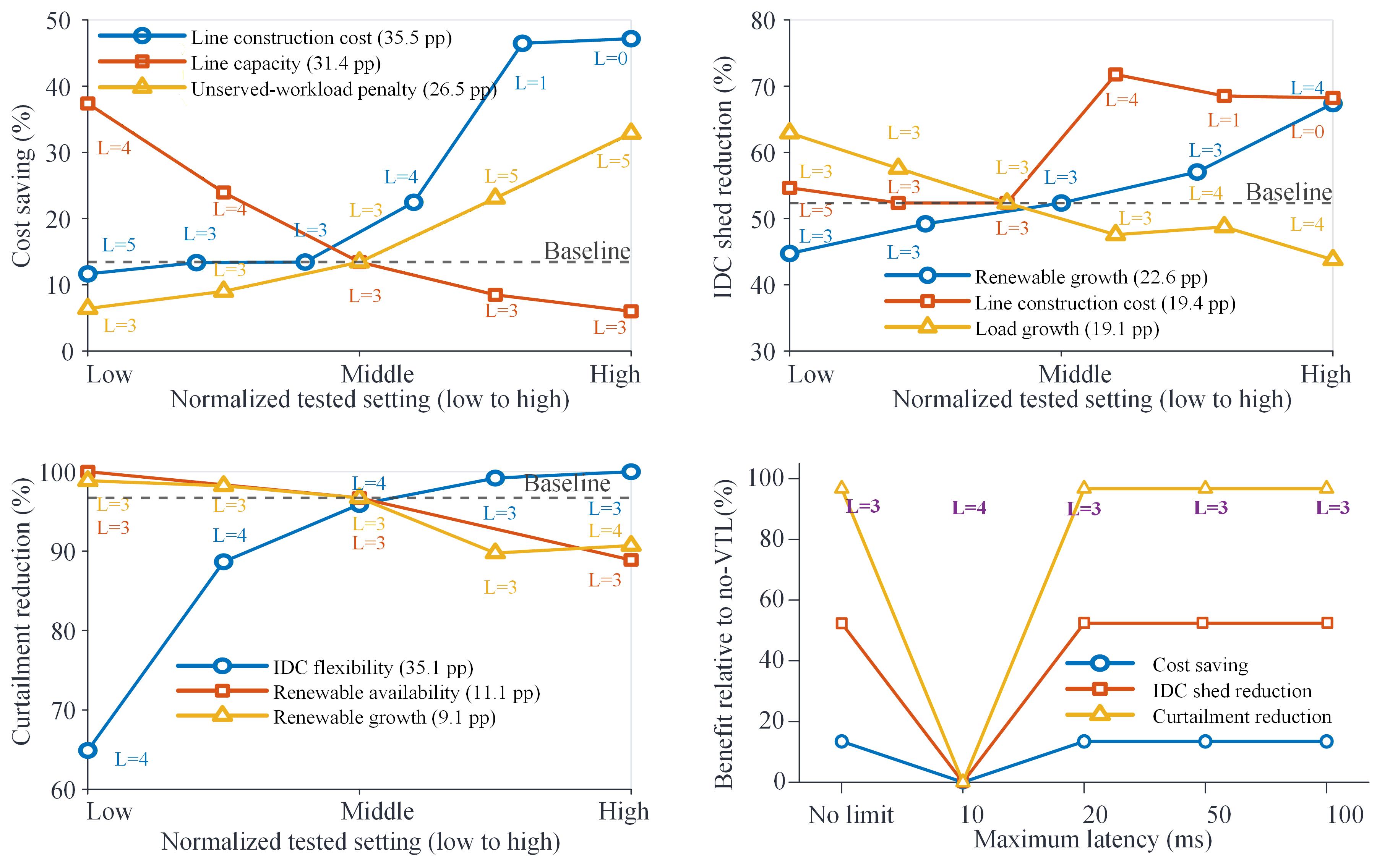}
\caption{Detailed response of VTL benefits to representative parameter variations: (a) cost savings under grid-side and economic parameters; (b) reduction in unserved DC load; (c) reduction in renewable curtailment; and (d) effects of maximum allowable communication latency relative to the no-VTL case. Dashed lines mark the baseline response; $L$ labels give the number of lines built by C3, and parenthetical values in the legends give the response span in percentage points. The latency panel represents discrete link availability rather than a continuous response.}
\label{fig:sensitivity_trends}
\end{figure}

Figure~\ref{fig:sensitivity_trends} shows the detailed responses for representative parameter variations.

Line-capacity and line-cost sweeps trigger discrete construction changes, marked by $L$. Bandwidth, data intensity, and communication cost have little effect over their tested ranges because communication cost and energy are small relative to the system objective and bandwidth is not binding in most runs. Maximum latency is different. Every modeled communication link has a latency of 20 ms (Appendix Table~\ref{tab:comm_links}). A 10-ms limit therefore removes all migration links and reduces the VTL benefit to zero, whereas limits of 20 ms or more retain all six links and restore the baseline response. This result is a discrete link-eligibility effect rather than a continuous latency response.

\subsection{Validation on the IEEE 118-bus system}

The IEEE 118-bus results further support the complementary relationship between VTL and physical transmission expansion. Over the five-year planning horizon, C3 reduces the total discounted cost from 44.25 billion yuan in C2 to 43.35 billion yuan, a reduction of 2.03\%. It also lowers the discounted investment cost by 31.22\% and reduces cumulative unserved DC load from 39.53 GWh to 7.34 GWh. C3 builds two candidate lines, compared with three in C2, while using 2.524 TWh of workload migration. Joint optimization also changes the investment schedule: the 69--75 corridor is commissioned one year earlier, the 89--90 corridor is commissioned one year later, and the 69--77 corridor selected in C2 is not constructed in C3. Compared with C0, C1 reduces unserved DC load by 78.71\%, but still leaves 93.12 GWh unserved over five years. These results suggest that workload migration can relieve network constraints and reduce selected transmission investments, but cannot fully replace physical reinforcement as DC demand grows. No renewable curtailment occurs in any of the four cases; thus, the benefits observed in this system arise primarily from lower investment requirements, reduced unserved DC load, and changes in generation dispatch rather than from additional renewable-energy utilization.

\section{Conclusion}
\label{sec:conclusion}

The results show where workload migration contributes to transmission planning and where it does not. Relative to build-only planning, the VTL-aware case reduces eight-year discounted cost by approximately 13.45\%, unserved DC load by approximately 52.4\%, and renewable curtailment by approximately 96.7\%, while reducing the optimized expansion from four lines to three. Under the fixed C2 construction plan, VTL lowers average loading by 3.23 percentage points and reduces fully loaded records by 1,583. These paired results link the benefit to nodal load redistribution rather than to a physical transfer of electricity.

The engineering implication is conditional. Migration is most valuable when transfer capacity is tight, construction is expensive, renewable output is difficult to deliver, conventional load grows rapidly, or a larger share of DC workload is movable. Yet C1's late-horizon unserved-load growth and the high-line-cost sensitivity cases show that VTL has a capacity boundary. It can relieve local constraints and improve which lines are built and when, but sustained DC-load growth still requires physical network expansion.

Future research will extend the proposed framework by coordinating temporal task deferral with spatial workload migration, thereby representing the broader spatiotemporal flexibility of DCs. Higher-resolution workload and communication data will also be incorporated to better characterize chronological variability, heterogeneous service requirements, and facility-level migration constraints. 

\appendix
\setcounter{table}{0}
\renewcommand{\thetable}{A\arabic{table}}
\renewcommand{\theHtable}{appendix.\arabic{table}}

\section*{Appendix}
\section{Complete model input data}

\begin{longtable}{@{}lrrrr@{\hspace{14pt}}lrrrr@{}}
\caption{Complete existing-line data. Reactance is in p.u. and the rating is in MW.}\label{tab:existing_lines}\\
\toprule
Line & From & To & $X_e$ & $\overline F_e$ & Line & From & To & $X_e$ & $\overline F_e$ \\
\midrule
\endfirsthead
\multicolumn{10}{l}{\textit{Table \thetable\ continued}}\\
\toprule
Line & From & To & $X_e$ & $\overline F_e$ & Line & From & To & $X_e$ & $\overline F_e$ \\
\midrule
\endhead
\bottomrule
\endlastfoot
L1 & 1 & 2 & 0.05917 & 102.00 & L11 & 6 & 11 & 0.19890 & 38.25 \\
L2 & 1 & 5 & 0.22304 & 59.50 & L12 & 6 & 12 & 0.25581 & 38.25 \\
L3 & 2 & 3 & 0.19797 & 46.75 & L13 & 6 & 13 & 0.13027 & 46.75 \\
L4 & 2 & 4 & 0.17632 & 59.50 & L14 & 7 & 8 & 0.17615 & 38.25 \\
L5 & 2 & 5 & 0.17388 & 59.50 & L15 & 7 & 9 & 0.11001 & 38.25 \\
L6 & 3 & 4 & 0.17103 & 46.75 & L16 & 9 & 10 & 0.08450 & 38.25 \\
L7 & 4 & 5 & 0.04211 & 76.50 & L17 & 9 & 14 & 0.27038 & 38.25 \\
L8 & 4 & 7 & 0.20912 & 46.75 & L18 & 10 & 11 & 0.19207 & 38.25 \\
L9 & 4 & 9 & 0.55618 & 38.25 & L19 & 12 & 13 & 0.19988 & 38.25 \\
L10 & 5 & 6 & 0.25202 & 46.75 & L20 & 13 & 14 & 0.34802 & 38.25 \\
\end{longtable}

\begin{longtable}{@{}lrrrrrr@{}}
\caption{Complete candidate-line data. Investment cost is undiscounted.}\label{tab:candidate_lines}\\
\toprule
Candidate & From & To & $X_c$ (p.u.) & $\overline F_c$ (MW) & Cost ($10^6$ yuan) & Lead time (year) \\
\midrule
\endfirsthead
\multicolumn{7}{l}{\textit{Table \thetable\ continued}}\\
\toprule
Candidate & From & To & $X_c$ (p.u.) & $\overline F_c$ (MW) & Cost ($10^6$ yuan) & Lead time (year) \\
\midrule
\endhead
\bottomrule
\endlastfoot
C\_2\_3\_1 & 2 & 3 & 0.19797 & 80 & 180 & 2 \\
C\_3\_4\_1 & 3 & 4 & 0.17103 & 80 & 180 & 2 \\
C\_4\_9\_1 & 4 & 9 & 0.55618 & 80 & 220 & 2 \\
C\_7\_8\_1 & 7 & 8 & 0.17615 & 80 & 180 & 2 \\
C\_7\_9\_1 & 7 & 9 & 0.11001 & 80 & 180 & 2 \\
C\_9\_14\_1 & 9 & 14 & 0.27038 & 80 & 160 & 2 \\
C\_2\_4\_1 & 2 & 4 & 0.17632 & 80 & 180 & 2 \\
\end{longtable}

\begin{longtable}{@{}lrrrrr@{}}
\caption{Complete conventional-generator data.}\label{tab:generator_data}\\
\toprule
Generator & Bus & $\underline P^G_g$ (MW) & $\overline P^G_g$ (MW) & $c^G_g$ (yuan/MWh) & $e_g$ (tCO$_2$/MWh) \\
\midrule
\endfirsthead
\multicolumn{6}{l}{\textit{Table \thetable\ continued}}\\
\toprule
Generator & Bus & $\underline P^G_g$ (MW) & $\overline P^G_g$ (MW) & $c^G_g$ (yuan/MWh) & $e_g$ (tCO$_2$/MWh) \\
\midrule
\endhead
\bottomrule
\endlastfoot
G1 (slack) & 1 & 0 & 332.4 & 320 & 0.75 \\
G2 & 2 & 0 & 140.0 & 360 & 0.80 \\
G3 & 3 & 0 & 100.0 & 450 & 0.60 \\
G4 & 6 & 0 & 100.0 & 500 & 0.70 \\
G5 & 8 & 0 & 100.0 & 520 & 0.70 \\
\end{longtable}

\begin{longtable}{@{}lrrrrrr@{}}
\caption{Bus base loads and complete DC parameters.}
\label{tab:bus_dc_data}\\
\toprule
DC & Bus & Compute-side capacity (MW)
& \(\eta_a\)
& \(r_a^{DC}\)
& \(\alpha_a\)
& \(r_a^{cap}\) \\
\midrule
\endfirsthead

\toprule
DC & Bus & Compute-side capacity (MW)
& \(\eta_a\)
& \(r_a^{DC}\)
& \(\alpha_a\)
& \(r_a^{cap}\) \\
\midrule
\endhead

\bottomrule
\endlastfoot

IDC\_load & 9 & 220 & 1.15 & 0.12 & 0.35 & 0.08 \\
IDC\_wind & 8 & 180 & 1.10 & 0.10 & 0.35 & 0.08 \\
IDC\_pv   & 3 & 180 & 1.10 & 0.10 & 0.35 & 0.08 \\

\end{longtable}

\begin{longtable}{@{}llrrr@{}}
\caption{Complete directed communication-link data.}\label{tab:comm_links}\\
\toprule
Origin & Destination & Bandwidth (GB/h) & Energy (kWh/GB) & Latency (ms) \\
\midrule
\endfirsthead
\multicolumn{5}{l}{\textit{Table \thetable\ continued}}\\
\toprule
Origin & Destination & Bandwidth (GB/h) & Energy (kWh/GB) & Latency (ms) \\
\midrule
\endhead
\bottomrule
\endlastfoot
IDC\_load & IDC\_wind & 6000 & 0.03 & 20 \\
IDC\_load & IDC\_pv & 6000 & 0.03 & 20 \\
IDC\_wind & IDC\_load & 4000 & 0.03 & 20 \\
IDC\_wind & IDC\_pv & 4000 & 0.03 & 20 \\
IDC\_pv & IDC\_load & 4000 & 0.03 & 20 \\
IDC\_pv & IDC\_wind & 4000 & 0.03 & 20 \\
\end{longtable}

\begin{longtable}{@{}lr@{\hspace{18pt}}lr@{}}
\caption{Representative-day weights used for annual scaling.}\label{tab:representative_weights}\\
\toprule
Scenario & $w_s$ (days/year) & Scenario & $w_s$ (days/year) \\
\midrule
\endfirsthead
\multicolumn{4}{l}{\textit{Table \thetable\ continued}}\\
\toprule
Scenario & $w_s$ (days/year) & Scenario & $w_s$ (days/year) \\
\midrule
\endhead
\bottomrule
\endlastfoot
spring\_normal & 30.638686 & autumn\_normal & 30.305657 \\
spring\_high\_renewable & 30.638686 & autumn\_high\_renewable & 30.305657 \\
spring\_peak\_load & 30.638686 & autumn\_peak\_load & 30.305657 \\
summer\_normal & 30.638686 & winter\_normal & 29.972628 \\
summer\_high\_renewable & 30.638686 & winter\_high\_renewable & 30.305657 \\
summer\_peak\_load & 30.638686 & winter\_peak\_load & 29.972628 \\
\midrule
\multicolumn{3}{r}{Total} & 365.000000 \\
\end{longtable}

\begin{longtable}{@{}llrrrrr@{}}
\caption{Raw optimized candidate-line construction schedules.}\label{tab:construction_schedule}\\
\toprule
Case & Candidate & Corridor & Start year & In-service year & Capacity (MW) & Cost ($10^6$ yuan) \\
\midrule
\endfirsthead
\multicolumn{7}{l}{\textit{Table \thetable\ continued}}\\
\toprule
Case & Candidate & Corridor & Start year & In-service year & Capacity (MW) & Cost ($10^6$ yuan) \\
\midrule
\endhead
\bottomrule
\endlastfoot
C2 & C\_2\_3\_1 & 2--3 & 2 & 4 & 80 & 180 \\
C2 & C\_4\_9\_1 & 4--9 & 3 & 5 & 80 & 220 \\
C2 & C\_7\_9\_1 & 7--9 & 1 & 3 & 80 & 180 \\
C2 & C\_2\_4\_1 & 2--4 & 4 & 6 & 80 & 180 \\
C3 & C\_2\_3\_1 & 2--3 & 1 & 3 & 80 & 180 \\
C3 & C\_4\_9\_1 & 4--9 & 2 & 4 & 80 & 220 \\
C3 & C\_2\_4\_1 & 2--4 & 3 & 5 & 80 & 180 \\
\end{longtable}

\section{Sensitivity settings and supplementary results}

\begin{longtable}{@{}p{0.18\textwidth}p{0.29\textwidth}p{0.43\textwidth}@{}}
\caption{Sensitivity parameters and tested settings.}\label{tab:sensitivity_ranges}\\
\toprule
Category & Parameter & Tested settings \\
\midrule
\endfirsthead
\multicolumn{3}{l}{\textit{Table \thetable\ continued}}\\
\toprule
Category & Parameter & Tested settings \\
\midrule
\endhead
\bottomrule
\endlastfoot
Grid & Line-capacity scale & 0.8, 1.0, 1.2 \\
Grid & Candidate-line cost scale & 0.5, 0.8, 1.0, 1.2, 1.5, 2.0 \\
Renewable & Annual renewable growth & 0, 0.03, 0.05, 0.08, 0.10 \\
Renewable & Renewable availability scale & 0.8, 1.0, 1.2 \\
Load/DC & Annual conventional-load growth & 0, 0.02, 0.04 \\
Load/DC & Migratable DC-workload ratio & 0.1, 0.2, 0.3, 0.35, 0.5, 0.7 \\
Load/DC & Unserved-load penalty & 5,000, 10,000, 20,000, 50,000, 100,000 yuan/MWh \\
Communication & Bandwidth scale & 0.25, 0.5, 1.0, 2.0, 4.0 \\
Communication & Data intensity & 10, 25, 50, 100, 200 GB/MWh \\
Communication & Data-transfer cost & 0, 0.001, 0.005, 0.01, 0.05 yuan/GB \\
Communication & Maximum allowable latency & No limit, 100, 50, 20, 10 ms \\
Economic & Carbon price & 0, 50, 100, 200, 300 yuan/tCO$_2$ \\
\end{longtable}

\begin{longtable}{@{}lrr@{}}
\caption{Overall congestion counts for C0--C3 over the eight-year horizon.}\label{tab:congestion_counts}\\
\toprule
Case & Loading $\geq95\%$ & Loading $=100\%$ \\
\midrule
\endfirsthead
\multicolumn{3}{l}{\textit{Table \thetable\ continued}}\\
\toprule
Case & Loading $\geq95\%$ & Loading $=100\%$ \\
\midrule
\endhead
\bottomrule
\endlastfoot
C0 & 42,492 & 35,685 \\
C1 & 44,076 & 35,257 \\
C2 & 37,737 & 30,299 \\
C3 & 38,139 & 30,340 \\
\end{longtable}

\begingroup
\setlength{\tabcolsep}{3.5pt}
\begin{longtable}{@{}lrrrrrr@{}}
\caption{Line-capacity and candidate-line investment-cost sensitivity. Costs are in 100 million yuan.}\label{tab:grid_sensitivity}\\
\toprule
Setting & C3 cost & Curtailment (MWh) & Unserved DC (MWh) & VTL migration ($10^4$ MWh) & C3 lines & C2--C3 saving \\
\midrule
\endfirsthead
\multicolumn{7}{l}{\textit{Table \thetable\ continued}}\\
\toprule
Setting & C3 cost & Curtailment (MWh) & Unserved DC (MWh) & VTL migration ($10^4$ MWh) & C3 lines & C2--C3 saving \\
\midrule
\endhead
\bottomrule
\endlastfoot
\multicolumn{7}{l}{\textit{Panel A: line-capacity scale}}\\
0.8 & 155.59 & 7,117.37 & 278,879.57 & 190.73 & 4 & 92.87 \\
1.0 & 119.87 & 1,855.38 & 80,418.27 & 198.02 & 3 & 18.62 \\
1.2 & 108.69 & 455.18 & 24,923.17 & 207.14 & 3 & 6.95 \\
\midrule
\multicolumn{7}{l}{\textit{Panel B: candidate-line cost scale}}\\
0.5 & 116.51 & 1,587.88 & 67,562.54 & 198.45 & 5 & 15.38 \\
0.8 & 118.77 & 1,855.38 & 80,418.27 & 198.02 & 3 & 18.32 \\
1.0 & 119.87 & 1,855.38 & 80,418.27 & 198.02 & 3 & 18.62 \\
1.2 & 123.87 & 1,587.88 & 90,026.69 & 209.83 & 4 & 35.89 \\
1.5 & 171.01 & 6,428.38 & 433,637.02 & 207.03 & 1 & 148.43 \\
2.0 & 171.79 & 6,685.89 & 454,459.63 & 208.45 & 0 & 153.35 \\
\end{longtable}
\endgroup

\begingroup
\setlength{\tabcolsep}{2.5pt}
\begin{longtable}{@{}lrrrrrrr@{}}
\caption{Renewable-growth and renewable-availability sensitivity. Costs are in 100 million yuan.}\label{tab:renewable_sensitivity}\\
\toprule
Setting & C2 cost & C3 cost & Saving & Curtail. reduction (MWh) & Unserved-DC reduction (MWh) & C2 lines & C3 lines \\
\midrule
\endfirsthead
\multicolumn{8}{l}{\textit{Table \thetable\ continued}}\\
\toprule
Setting & C2 cost & C3 cost & Saving & Curtail. reduction (MWh) & Unserved-DC reduction (MWh) & C2 lines & C3 lines \\
\midrule
\endhead
\bottomrule
\endlastfoot
\multicolumn{8}{l}{\textit{Panel A: annual renewable-growth rate}}\\
0 & 143.94 & 126.13 & 17.81 & 28,282.33 & 83,486.74 & 4 & 3 \\
0.03 & 140.60 & 122.36 & 18.24 & 39,160.66 & 86,338.79 & 4 & 3 \\
0.05 & 138.50 & 119.87 & 18.62 & 54,518.76 & 88,384.24 & 4 & 3 \\
0.08 & 135.68 & 116.30 & 19.37 & 96,872.52 & 91,424.29 & 4 & 3 \\
0.10 & 134.07 & 113.59 & 20.48 & 151,864.37 & 104,649.34 & 4 & 4 \\
\midrule
\multicolumn{8}{l}{\textit{Panel B: renewable-availability scale}}\\
0.8 & 144.01 & 126.12 & 17.88 & 9,493.67 & 86,252.95 & 4 & 3 \\
1.0 & 138.50 & 119.87 & 18.62 & 54,518.76 & 88,384.24 & 4 & 3 \\
1.2 & 134.66 & 114.70 & 19.96 & 155,446.79 & 90,497.40 & 4 & 3 \\
\end{longtable}
\endgroup

\begingroup
\setlength{\tabcolsep}{2.5pt}
\begin{longtable}{@{}rrrrrrrr@{}}
\caption{Complete carbon-price sensitivity. Costs and savings are in 100 million yuan.}\label{tab:carbon_sensitivity}\\
\toprule
Carbon price & C2 cost & C3 cost & Saving & Curtail. reduction (MWh) & Unserved-DC reduction (MWh) & C2 lines & C3 lines \\
\midrule
\endfirsthead
\multicolumn{8}{l}{\textit{Table \thetable\ continued}}\\
\toprule
Carbon price & C2 cost & C3 cost & Saving & Curtail. reduction (MWh) & Unserved-DC reduction (MWh) & C2 lines & C3 lines \\
\midrule
\endhead
\bottomrule
\endlastfoot
0 & 121.86 & 103.23 & 18.64 & 54,518.76 & 88,384.24 & 4 & 3 \\
50 & 130.18 & 111.55 & 18.63 & 54,518.76 & 88,384.24 & 4 & 3 \\
100 & 138.50 & 119.87 & 18.62 & 54,518.76 & 88,384.24 & 4 & 3 \\
200 & 155.13 & 136.52 & 18.61 & 54,518.76 & 88,384.24 & 4 & 3 \\
300 & 171.76 & 153.16 & 18.60 & 54,518.76 & 88,384.24 & 4 & 3 \\
\end{longtable}
\endgroup

\bibliographystyle{unsrt}
\bibliography{references}

\end{document}